From Edward.Shuryak@cern.ch Mon Jun 25 04:20:43 2001
Date: Mon, 25 Jun 2001 10:18:02 +0200 (MET DST)
From: Edward Shuryak <Edward.Shuryak@cern.ch>

\documentstyle[twocolumn,prc,aps,epsfig]{revtex}
\input epsf
\newcommand{\be}{\begin{eqnarray}}
\newcommand{\ee}{\end{eqnarray}}
\newcommand{\no}{\nonumber \\}
 \newcommand{\gsim}{\mathrel{\hbox{\rlap{\lower.55ex \hbox {$\sim$}}
                   \kern-.3em \raise.4ex \hbox{$>$}}}}
\newcommand{\lsim}{\mathrel{\hbox{\rlap{\lower.55ex \hbox {$\sim$}}
                   \kern-.3em \raise.4ex \hbox{$<$}}}}
\def\calO{{\cal O}}
\def\calB{{\cal B}}
\newcommand{\e}{{\mbox{e}}}
\def\del{\partial}
\def\vr{{\vec r}}
\def\vk{{\vec k}}
\def\vq{{\vec q}}
\def\vp{{\vec p}}
\def\vP{{\vec P}}
\def\vt{{\vec \tau}}
\def\vs{{\vec \sigma}}
\def\vJ{{\vec J}}
\def\vB{{\vec B}}
\def\hatr{{\hat r}}
\def\hatk{{\hat k}}
\def\roughly#1{\mathrel{\raise.3ex\hbox{$#1$\kern-.75em%
\lower1ex\hbox{$\sim$}}}}
\def\lsim{\roughly<}
\def\gsim{\roughly>}
\def\fm{{\mbox{fm}}}
\def\vx{{\vec x}}
\def\EM{{\rm EM}}
\def\barp{{\bar p}}
\def\zz{{z \bar z}}
\def\mus{{\cal M}_s}
\def\abs#1{{\left| #1 \right|}}
\def\ve{{\vec \epsilon}}
\def\nlo#1{{\mbox{N$^{#1}$LO}}}
\def\MS{{\mbox{M1V}}}
\def\mut{{\mbox{M1S}}}
\def\Qt{{\mbox{E2S}}}
\def\rM{{\cal R}_{\rm M1}}\def\rE{{\cal R}_{\rm E2}}
\def\la{{\Big<}}
\def\ra{{\Big>}}
\def\J#1#2#3#4{ {#1} {\bf #2} (#4) {#3}. }
\def\PRL{Phys. Rev. Lett.}
\def\PL{Phys. Lett.}
\def\PLB{Phys. Lett. B}
\def\NP{Nucl. Phys.}
\def\NPA{Nucl. Phys. A}
\def\NPB{Nucl. Phys. B}
\def\PR{Phys. Rev.}
\def\PRC{Phys. Rev. C}
\renewcommand{\thefootnote}{\arabic{footnote}}
\begin{document}

\twocolumn[\hsize\textwidth\columnwidth\hsize\csname @twocolumnfalse\endcsname

\title {
 The Instanton/Sphaleron Mechanism of Prompt Gluon Production\\ in
High Energy Heavy Ion Collisions at RHIC}
\author {  Edward V. Shuryak \\
 Department of Physics and Astronomy\\ State University of New York,
     Stony Brook, NY 11794-3800
}
\date{\today}
\maketitle
\begin{abstract}
We argue that if the growing part of hadron-hadron cross section
(described phenomenologically by the supercritical soft Pomeron)
is due to instanton/sphaleron mechanism, one should find certain
qualitative features of the produced cluster which differ from
the usual string fragmentation. Furthermore,
we suggest that this mechanism
 should be even more important for  heavy ion collisions
in the RHIC energy domain. 
Large number of parton-parton collisions should result in hundreds
of  produced sphaleron-like
gluomagnetic clusters per unit rapidity. Unlike perturbative gluons
(or mini-jets), these {\em classically unstable}
objects promptly decay into several gluons and quarks in 
mini-explosions, leading to very rapid entropy generation.
This may help to explain why the QGP seem to be produced at RHIC so early.
We further argue that this mechanism cannot be important
at  higher energies (LHC), where  perturbative description should
apply.
\end{abstract}
\vspace{0.1in}
 ]
\newpage
1.At high energies $s>10^3 \, GeV^2$  hadronic cross sections ($\bar p p,pp,\pi p, Kp$,
   $\gamma N$ and even $\gamma\gamma$)  
 slowly grow  with the collision
energy s. This behavior can be  parameterised
by a Regge pole, the so called {\em soft Pomeron} (see e.g. \cite{DL}).
In this Letter we will not address very high s and therefore use only the logarithmic 
fit
\be \sigma_{hh'}(s)= \sigma_{hh'}(s_0)+ log (s/s_0) X_{hh'}\Delta +...
\ee 
 ignoring  both the question of whether
 it is indeed a Regge pole,
as well as other Reggions leading to contribution decreasing with
energy. For estimates below we
 use  parameters from the latest  Particle Data Group fits \cite{pdg},
which give  the ``Pomeron intercept'' 
 and a constant equal to
$pp,\bar p p$ collisions,
$ \Delta=  \alpha(0)-1=  0.093(2), \, X_{NN}= 18.951(27)\ mb. $

\begin{figure}[t]

  \epsfxsize=1.3in
  \centerline{\epsffile{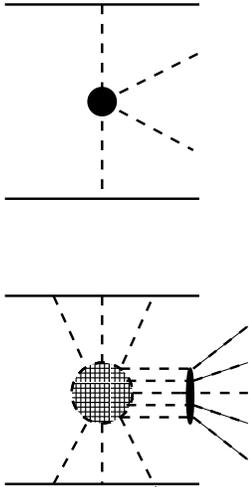}}

  \caption[]
  {
   \label{fig}
(a) A typical inelastic perturbative process: two t-channel gluons collide,
producing
a pair of gluons; (b) Instanton-induced inelastic process incorporate
collisions
of multiple t-channel gluons with the  instanton (the shaded circle),
resulting in multi-gluon production. The intermediate stage of the
process,
indicated by
the horizontal dashed lines, corresponds to a time when outgoing glue
is
in the form of coherent field configuration - the $sphaleron$.
  }

\end{figure}

A qualitative  difference between constant and logarithmically growing parts of the
cross section will be emphasised. The former can be explained by prompt color
$exchanges$, as suggested by Low and Nussinov \cite{LN} long ago.
The $growing$ part of the cross section
 cannot be generated by t-channel vector exchanges and is
 associated with  prompt (prior to formation of strings)  production
of some objects,
with  log(s) coming  from  longitudinal phase space.
Perturbative QCD  describes  {\em gluon} production, by processes like the one
shown in Fig.\ref{fig}(a), which can be 
iterated in the t-channel
in ladder-type fashion resulting in  a BFKL pole
 \cite{BFKL}.
Although its intercept
 is much larger than $\Delta$ mentioned, 
 it is
 consistent with much stronger growth seen in hard processes at
HERA: thus it is often called
a ``hard pomeron''. 

 The physical origin of growing cross section remains an outstanding open
problem: 
neither the  perturbative resummations
nor existing
non-perturbative models are really quantitative.
It is hardly surprising, since the  scale at which  soft Pomeron
operates
(as seen e.g. from the Pomeron slope $\alpha'(0)\approx 1/(2 \, GeV)^2$)
is  the  semi-hard or
 {\em ``substructure scale''} 
   $Q^2 \sim \, 1-2 GeV^2$, which is notoriously difficult for theorists
because it is
simultaneously
the {\em lower} boundary of pQCD (serving therefore as the cut-off
$p_{min}$
already mentioned), 
as well as the {\em upper} 
 boundary of 
low energy effective approaches like chiral Lagrangians
or Nambu-Jona-Lasinio model.
At the same time, 
a number of  objects/phenomena are naturally ascribed to  
  this scale:   ``constituent quarks'', flux tubes
(or QCD strings) and their junctions, to name a few.
We do not have a quantitative description of flux tubes (other than
lattice QCD), but
constituent quarks and related issues can be
well understood in an 
{\em instanton liquid model}, see review \cite{SS_98}.
Its primary parameters  are the number density of instantons
(plus anti-instantons) and their average size, determined long ago 
\cite{Shu_82},
from QCD phenomenology to be
$n \approx 1$~fm$^{-4}$ and small average size of $\bar\rho\approx
1/3$ fm leading to vacuum diluteness $n\rho^4\sim 10^{-2}$.
Amazingly, with those two numbers one can get truly quantitative
description of  correlators, form-factors and other hadronic
parameters\footnote{For recent example
see \cite{SS_00} where
vector and axial correlators obtained from the $\tau$ decays
 are explained literally within their  error bars, or withing few percent
accuracy.}. 
 
Application of the instanton-induced dynamics
to  high energy hadronic collisions have been suggested recently \cite{S_pom_glueballs,pom_inst}. One
 important precursor has been the Kharzeev-Levin work \cite{KL}
in which contribution to $\Delta$  of scalar colorless states --
the sigma meson and the scalar glueball -- has been non-perturbatively
 evaluated. 
 Two last works in \cite{pom_inst} have benefited from
  deep  insights obtained a decade ago in
studies of instanton-induced processes in electroweak theory,
see \cite{weakinst} and references therein. 
In these works
 the growing part of the hh cross sections is due to prompt 
multi-gluon
production via instantons, or more accurately, via $colored$ gluonic clusters
called sphalerons,
see  Fig.\ref{fig}(b).

 Among qualitative features of this theory
is the explanation of why no odderon appears (instantons are SU(2)
objects, in which quarks and antiquarks are not really distinct),
an explanation of the small power $\Delta$ (it is proportional to
``instanton diluteness parameter'' $n\rho^4$ mentioned above), 
the small size of the soft Pomeron
(governed simply by small size of instantons mentioned above,
$\rho\sim 1/3 \, fm$). 
Although instanton-induced amplitudes are proportional to small
``diluteness''
factor, there is {\em no extra penalty for production of new 
gluons}: thus one should  expect instanton effects to beat perturbative
amplitudes
of sufficiently high order. This generic idea is also behind the
present
work, dealing with prompt multi-gluon production.


 Technical description of the process is split into two stages. 
The first (at which one evaluates the cross section) 
is the motion {\em under the barrier},  described by 
Euclidean paths approximated by  
instantons. Their
interaction with the high energy
colliding partons results in some energy deposition and subsequent 
real motion  {\em above the barrier}. At this second stage
the action is real, and the factor  $|exp(iS)|=1$, so it
 does not affect the cross section and is need only to detail
the final state.
The relevant Minkowski paths start with 
configurations  close to
QCD analogs of electroweak {\em sphalerons} \cite{Manton}, 
 static spherically-symmetric clusters
 of gluomagnetic field\footnote{
 Those can be 
obtained from known electroweak solutions
 in the limit of infinitely large Higgs self-coupling.}. Their mass in
QCD is also determined by the isnatnton size
\be M_{sph}\approx {30 \over g^2(\rho) \rho}\sim  2.5 \, GeV \ee
Since those field configurations
are close  to  classically unstable saddle point
at the top of the barrier, they roll downhill and develop gluoelectric
fields.  When both become  weak enough,
solution can be decomposed  into perturbative    
gluons. This part of the process can also be studied directly from
classical
Yang-Mills equation: for electroweak sphalerons it has been done in
Refs\cite{sphaleron_decay}, calculation
for its QCD version is in progress \cite{CS}.
While rolling,  the configurations tend to forget the initial
imperfections (such as a non-spherical shapes) since there is
only one basic instability path downward:  so
 the resulting fields
should be  nearly perfect spherical expanding shells.
 Electroweak sphalerons  decay into approximately 51
W,Z,H quanta, of which only about 10\%  are Higgses, which carry only
4\% of energy. Ignoring those, one can make crude {\em tentative 
estimate} of mean gluon multiplicity per  sphaleron decay,
by simple re-scaling of the coupling constants 
\be 
<N_g>\approx <N_{W,Z}> {g_{electroweak}^2 \over g_{QCD}(\rho)^2}\sim 3-4
\ee
 The spectrum  (also derived from a
solution \cite{sphaleron_decay,CS})  has a wide maximum and can
roughly be approximated by
 thermal one, with a temperature of
about $T_0\sim 300 MeV$, see  \cite{CS}.

2.The first points we would like to make in this Letter are some suggestions of
how one can experimentally test this scenario, by some qualitative effects. 

Note that the fate of the produced sphalerons is different in hh and AA
collisions: in the former case they decay
 in the confining  vacuum, in the latter into a deconfined media (see below).
Some fraction of produced clusters have net zero color and can directly form
 glueballs,  with $J^P=0^+,0^-$ \footnote{But not a channel with e.g.
  $2^+$ quantum numbers,
 which does not classically couple to instantons .}. The scalar isoscalar channel    
 has been considered first in 
\cite{KL}:
 however, it can only account for a fraction\footnote{In terms of pomeron intersept, it is $\Delta_{0^+}\approx 0.05$  for scalar glueball and sigma together \cite{S_pom_glueballs} , while (including shadowing) the experimental
total is $\approx 0.16$ \cite{COS}.} of prompt production. Most of the promptly 
produced gluon clusters have non-zero net color, and
 the thus have to be
 connected by
 color flux tubes to other partons. 

This clearly makes
their direct observation difficult, but not hopeless: we briefly describe two
particular ideas of how it can be done. 
The scalar glueball candidate $f_0(1700)$ decay into $\eta\eta,\bar K K$ and only a little bit
 into
$\pi\pi$. We do not yet have experimental pseudo-scalar glueball candidate,
 while
lattice predicts it to be right at the mass of the sphaleron (2).
However, as noticed by Bjorken \cite{BJ}, the  $\eta_c$ decay
has  three
distinct 3-meson modes, $KK\pi, \eta\pi\pi, \eta'\pi\pi$,
 with about 0.05 branching each: those fit well to the idea that they come
directly
from the 't Hooft instanton-induced Lagrangian  $\bar u u \bar d d \bar s s$.
Presumably the instanton-induced 
decays modes of the $0^-$ glueball should prefer the same 3 channels. 

 One may  speculate further, and suggest that scalar
(pseudo-scalar) projections of the sphaleron 
may still follow the same
scalar (pseudo-scalar) glueball decay pattern,  even while
 the total color is non-zero. The pattern of enhanced production
 of 
 $\eta',\eta,K$ via strange part of 't Hooft Lagrangian leads to a specific
frature of the final state. Indeed, when
 $\eta',\eta,K_S$ decay  into  5,3,2 pions, respectively, 
 all of them are produced much later than the average pion production time. 
They are different from others in one important aspect: they {\em do not participate in Bose-Einstein (or HBT)  correlations}. 
Its strength is traditionally
expressed in terms of the so called $\lambda_{HBT}=(1-f)^2$, where
$f$ is the fraction of pions coming from long-lived\footnote{Defined
relative to $\hbar/\Delta E$ where $\Delta E$ is the
 energy resolution of the detector.} sources.
In minimal bias pp, or heavy ion collisions
with $any$ multiplicity, or in the $e^+e^-$ reactions
 the usual value   $\lambda_{HBT}\approx 0.5$. However
  for high multiplicity  $\bar p p$ collisions
 experiments show  that the intensity of the correlations  
decreases substantially, 
to only
$\lambda_{HBT}\approx  0.2$. As far as we know, this effect has not been 
explaned: 
see discussion of data and proposed suggestions  in \cite{lambda}.
Although clearly much more studies are needed, it may indicate
that  promptly produced hadrons have  an
origin other than the usual string fragmentation.

Another  possible approach is based on the  (so far ignored) topological
  properties
of instantons/sphalerons. Roughly speaking, each sphaleron 
has an option to roll down in two directions,
selecting two possible orientations of its gluoelectric field relative to gluomagnetic one.
Parity is of
  course conserved in QCD: but on the  $event-by-event$ basis 
large fluctuations may appear in P- and CP-odd
kinematical observables specified in \cite{CP}.

3.We now turn to heavy ion collisions.
Recent experiments at Relativistic
Heavy Ion Collider (RHIC) at Brookhaven National Laboratory, taken
during its first run in summer 2000 and reported recently at  Quark
Matter 2001 conference
 \cite{QM01},
 have shown that
 heavy ions collisions (AA) at highest energies 
significantly differ 
$both$ from the hh collisions 
 and  the AA collisions at lower (SPS/AGS) energies.
Many features of these data are quite consistent with
 the Quark-Gluon Plasma (QGP) (or Little Bang) scenario 
\cite{Shu_80}, in which entropy is produced promptly and
subsequent expansion is close to adiabatic expansion of equilibrated
hot medium.

Already the very first multiplicity measurements
 reported by PHOBOS collaboration  \cite{PHOBOS}
 have shown that particle production per participant
  nucleon is no longer constant, 
as was the case at lower (SPS/AGS) energies, but grows more rapidly.
This behaviour may  be due to long-anticipated {\em pQCD}
processes, leading to perturbative production of new partons. Unlike
high $p_t$ processes, those are
(directly undetectable)  {\em ``mini-jets''}. Their
 production and decay was discussed
 in Refs \cite{minijets}, and also used in
 widely used event generator HIJING  \cite{HIJING}. 
Its crucial parameter is the {\em cutoff scale} $p_{min}= 1.5-2  \,GeV$: if
fitted from pp data to be, it leads to predicted mini-jet multiplicity
$dN_g/dy\sim 200$ for central AuAu collisions at $\sqrt{s}=130 \,
AGeV$. If those fragment independently into hadrons, and are supplemented
  by ``soft'' string-decay component, the predicted total multiplicity
was found to be in good agreement with the
first RHIC multiplicity data.  Because  partons
interact perturbatively, 
with their scattering  and radiation being strongly peaked at small angles,
  their equilibration is expected to be  relatively long
  \cite{equilibr}.

However, next set of  RHIC data reported in  \cite{QM01} 
have provided  serious arguments {\em against} the mini-jet  
scenario, and point toward quite rapid entropy production rate and
early QGP formation.

(i) If most of secondaries come from independent mini-jets fragmentation, there would be
no {\em collective phenomena}
such as transverse flow 
related with the QGP pressure. However, 
such effects are very strong at RHIC. In particular,
STAR collaboration have observed very robust
{\em elliptic flow} 
 \cite{elliptic_STAR}, which is in perfect agreement with
predictions of hydrodynamical model
 \cite{elliptic_hydro}
 assuming equilibrated QGP with its
full pressure $p\approx \epsilon/3$ above the QCD phase
transition. This agreement persists to rather peripheral
collisions, in which the overlap almond-shaped region of two nuclei
is only a couple fm thick. STAR and PHENIX data on spectra of
identified
particles, especially $p,\bar p$, indicate spectacular radial expansion,
also
in agreement with hydro calculations  
 \cite{elliptic_hydro}.

 (ii) Spectra of hadrons at large $p_t$, especially the $\pi^0$
spectra
 from PHENIX,
 agree well with HIJING for peripheral collisions, but  show 
much smaller yields
for central ones, with rather
different spectra both in shape and composition.
Moreover, those agree weel with hydro predictions which had been
established at low $p_t$ previously.
 It means that not only  long-anticipated
{\em  ``jet quenching''}  is observed, it seems to be as
large
 as it can possibly be\footnote{
 Jets originating
from the surface outward is very difficult to quench, and thus the  
   suppression factor of about $\sim 1/10$ is difficult to decrease further,
whatever happens in dense matter.
Counting from expected Cronin effect
(which in pA collisions is about factor 2 at $p_t$ in question), the observed
suppression is not far from such number.}.
 For that to happen, the outgoing high-$p_t$
 jets  should propagate through matter with parton population much
   larger than the abovementioned
minijet density predicted by pQCD (HIJING).

(iii) Curious interplay between collective and jet effects have
also been studied by STAR collaboration, in form 
of elliptic asymmetry parameter $v_2(p_t)$. At large 
transverse momenta $p_t>2\, GeV$ the data
behave according to
predictions of jet quenching model \cite{GW},
  indicating gluon
 multiplicity several times larger than HIJING prediction. Moreover,
the result is
  in fact consistent with
the $maximal$ possible value  evaluated from  the final 
  entropy at freeze-out, $(dN/dy)_\pi\sim 1000$.

 In this Letter we propose a non-perturbative
solution to this puzzle. But before we come to it, let us also mention
its alternative:  {\em significantly 
 lower cutoff scale}  in excited matter, as compared to $p_{min}=1.5-2\,GeV$
fitted from the pp data. It may lead to larger perturbative cross sections,
both due to smaller momenta transfer and larger coupling constant.
As  argued over the years (see e.g. one of the talks
\cite{Shur_cutoff}), the QGP is a new phase of QCD which is
 {\em qualitatively different} from the QCD
vacuum:
   therefore the  cut-offs of pQCD may  have entirely different values
and be determined by different phenomena.
 Furthermore, since QGP is a plasma-like phase which screens itself
perturbatively \cite{Shu_80}, one may think of a   cut-offs to   
 be determined {\em self-consistently} from resummation of perturbative
effects. These ideas known as {\em self-screening} or
{\em QGP saturation} were discussed in Refs.
 \cite{selfscreening}. Although the scale in question grows with 
temperature or density, {\em just above $T_c$} it may actually be  
{\em smaller} than the value 1.5-2 GeV observed in  vacuum. 
Its direct experimental manifestation may be
deformation of dilepton spectra, whcih can be well
described by decreasing
``duality scale''  \cite{RW}.

4.In order to specify the magnitude of new production mechanism 
 one can study
dependence on the impact parameter $b$. This dependence
of the pseudo-rapidity density at mid-rapidity,
 measured 
at RHIC \cite{QM01} can be very accurately described by
simple parameterization  \cite{KN}
\be
{dn_{AuAu}(\eta=0,b)/d\eta \over dn_{NN}(\eta=0)/d\eta}={(1-x)\over 2}N_{part}(b)+xN_{coll}(b)
\ee
where  the average number of
participants
$N_{part}(b)$ and NN collisions $N_{coll}(b)$   are calculated
in standard  Glauber model.

 The key is new (b-independent) parameter  $x(s)$, 
defining which  fraction of NN collisions  scales differently in AA.
  We  propose to identify 
$x(s)$  with the {\em growing part
of the NN cross section} discussed above, namely
\be x(s)=\Delta {X_{NN}\over  \sigma_{hh'}(s_0)} log(s/s_0) \ee
Note that two phenomenological values fitted at two RHIC energies
 $x(\sqrt{s}=56\, Gev)=0.05\pm0.03,x(\sqrt{s}=130\, Gev)=0.09\pm0.03$ \cite{KN}
are both well reproduced if one selects the
threshold 
value at $s_0=1000\, GeV^2$, the position of the NN cross 
section $minimum$. Furthermore, because
 this $s_0$  is above the highest SPS 
 energy, it explains why this component  has not been seen before.
This identification is due to the picture of  prompt production
of some objects -- minijets or sphalerons --
 in partonic collisions \footnote{
The $const(s)$ part of the cross section, which 
is associated with color exchanges, should scale as the
number of participants because, no matter how many exchanges took place,
each outgoing parton pulls out {\em only one} color flux tube
per quark (or 2 per gluon).}.

Partons which participate in such interaction  should
 be appropriately normalized at the scale 
$\mu^2\sim 1-2 \, GeV^2$.
Constituent  quark models of 60's would 
count only 
them, so  $N^{baryons}_{p}=3$. 
 Using  parton densities derived 
from structure functions one finds that (at  scale under discussion)
 sea  can be
neglected, but gluons do not. With
significant uncertainties,  for RHIC energy one can
integrate structure functions for $x>0.01$ and get 
roughly additional  3 gluons, leading to
 $N^{N}_{p}\approx 6$ \footnote{Detailed evaluation of semi-hard
partonic cross sections from (i) the growing part
of all hadron-hadron scattering cross sections, (ii) elastic amplitudes and 
(iii) p,$\pi,\gamma$ structure functions will be reported elsewhere \cite{COS}.}.

The inelastic hh' cross section can be schematically written in a simple multiplicative form
\be 
\sigma_{hh}=N^{h}_{p}N^{h'}_{p} (\sigma^0_{pp}+\sigma^1_{pp} log s +O( log^2(s))
\ee
For
simplicity
of presentation, we ignore the difference between $qq,\bar q q, qg,gg$ 
cases, as well as possible dependence on quark flavor.
Here
 $N^{h}_{p}$ is the number of partons per
hadron
and $\sigma^0_{pp},\sigma^1_{pp}$ are the  parton-parton cross sections,
without and with prompt production. In what follows we
ignore the former and only concentrate on the latter part, 
normalizing it to the observed soft Pomeron
growth $\sigma^1_{pp}=X_{NN}/(N_p^N)^2$.
Bypassing dynamical calculations  \cite{pom_inst}, we 
then estimate the 
probability of the sphaleron production directly from data,
by  assuming it to be the dominant process behind  
 the logarithmic growth of the cross section.

 It means that in mean parton-parton
collision,
the cross section per rapidity of prompt production\footnote{Note a {\em surprisingly small},
 factor 1/100,  compared to 
  geometric cross section $\pi \rho^2$.
In  instanton-based theory
it originates directly
from the first power \cite{pom_inst}
of  instanton diluteness of such magnitude.
}
\be 
{d\sigma_{prompt} \over dy} = {X_{NN}\Delta \over ( N^{N}_{p})^2}\sim
0.005\, fm^2,
\ee
 Now we evaluate the total number of parton-parton
collisions in $central$ AA collision. Unlike the total cross section, it is not just
a multiplicative expression:  nuclear geometry 
leads to
\be 
N_{pp \ collisions}(AA)\sim (A*N^N_p)^{4/3}\sim 10^4
\ee
where in numerical estimate
we have used  $A=200$. 
Assuming simple factorization of the cross sections\footnote{
Note that we are still very far from unitarity
constraints. Inside the tube with instanton radius $\pi\rho^2$ we find about
.67 partons in a nucleon and 3.6 in Au: so even factorized cross section lead
to interaction probablity of only 1/200 and 1/10 respectively, much less than 1. It does not mean however that
 factorization is accurate: we use it only as an estimate.
For instanton processes presence of extra partons lead to
extra factors -- Wilson lines -- in the amplitude, but averaging  over
 instanton collective variables (such as color orientations)
may upset factorization. This question deserves  quantitative study.
Note also, that partons found at the same position in transverse plane
 most likely come from different nucleons, so
position and color correlations between them are likely to be small:
so no assumptions about wave functions are probably needed here.
}.
Combining these two simple
ingredients we now estimate
the total density of ``promptly-produced objects''
(mini-jet pairs or sphalerons) in AA
collisions per unit rapidity
\be 
{dN_{prompt} \over dy} = ({X_{NN} \Delta \over \pi \rho^2}) {A^{4/3} 
\over (N_{p}^N)^{2/3}} \sim 200
\ee
 Presumably one can still treat these objects as produced independently, since 
 the number of available
cells in the transverse
plane  $N_{cells}=(R/\rho)^2\approx 400$ is still larger than 
this $maximal$  sphaleron number. 

The number of ``promptly produced objects'' 
estimated above is rather close to
 mini jet-production\footnote{As should be expected, since minijet models
fit the pp cross section as well.} calculated with 
HIJING \cite{HIJING}.
Furthermore,
multiplying it by the transverse energy (2/3)$m_{sph}$, 
we find that our prompt production should 
result in roughly $dE_t/dy\sim 400 \, GeV$ of transverse energy,
again comparable to HIJING predictions. 
So a critical reader  may ask whether
 actually anything  has been gained, by substituting
one hypothetical mechanisms of prompt production --
the mini-jet  scenario --  by another one, based on instantons/sphalerons.
Indeed,
provided both are similarly normalized to  growing
part of the pp collisions and then scaled to  AA case, 
we get about the same number of semi-hard events and the sama excitation energy.

Our first (theoretical) 
answer is that the suggested scenario suggests an explanation to the semi-hard scale involved,  derived from the well known vacuum instanton parameters, while in pQCD the cutoff should be just guessed or fitted. 
 Furthermore, it
implies
detailed microscopic knowledge of the specific gluon field
configuration involved, 
not just estimate of a number of gluons produced. 

The second (pragmatic)  answer is that these two scenarios differ  
significantly in the {\em amount of the
entropy produced}.
The minijets
 are just plane waves:  they are classically stable and weakly interacting.
The sphalerons are unstable, a kind of
resonances existing already at classical level.
They  explode into spherical
expanding shells
of strong field, which  rapidly sweep the whole volume
and may convert it into Quark-Gluon Plasma, in which
the charge is screened 
rather than confined
 \cite{Shu_80}. The ``initial temperature'' of gluons produced
from sphaleron decay $T\sim 300 \, Mev$
indicated above is definitely above the critical value.
 Most important, the produced entropy is  several times
 larger than for minjets, as
recent RHIC data seem to indicate. 

In heavy ion collisions at RHIC
the QGP is supposed to exist at RHIC for several fm/c,
 much longer than
the sphaleron  lifetime
$\tau_{sph} \sim 1/\rho$. 
If so, partons produced
do not hadronize immediately (as for hh collisions)
but decay into 3-4 gluons,
plus 0-6 quarks\footnote{ Although in QGP  there are no
quark condensates and one may think that all 6 
't Hooft $\bar u u \bar d d \bar s s$ are produced, it is not nacessarily so since they could still be from the
initial vacuum. Evaluation of probablities for
each quark multiplicity we hope to report elsewhere.} and start real equilibration.

Phenomenologically,
comparing $dN_{sphalerons}/dy\approx 200$ sphalerons
to  $dN_{gluons}/dy\approx 1000$ one sees that  about 5 partons/sphaleron
 would produce the right amount of entropy.
that about 5 partons/sphaleron would do the job, which is concievable. 
 In order to test
  the conjectured mechanism experimentally, one may try to infer
  gluon/quark ratio at early time from dilepton production. Another possibility is to look at event-by-event fluctuations following from clustering
at the production stage. 

5.Finally, let us briefly discuss
what  we predict should happen at much higher collision energy, e.g. 
at CERN LHC? At what partonic scale the main processes
will  be stabilized?
A plausible answer
 suggested in Ref. \cite{saturation,McLV} is that
 high parton density will generate its own
 $saturation$   scale, estimated for
 LHC to be about 
 $\mu^2\sim 10 \,GeV^2$.

 If so,
   the instanton/sphaleron mechanism  described above 
can no longer be important.  The reason for that is extremely sharp
dependence of the instanton effects on the scale involved, originating
in semiclassical action
$exp(-S)\sim (\Lambda_{QCD}/\mu)^{(11/3)N_c-(2/3)N_f}$.
 Therefore, if going
from
RHIC to LHC we  change $\mu$  by  factor 3, 
 the sphaleron production is expected to drop  by 
 3-4 orders of magnitude, becoming much less than its pQCD background.

{\bf Acknowledgments.}
It is a pleasure to thank G.Carter,D.Kharzeev,L.McLerran, M Gyulassy
and especially  I.Zahed for valuable discussions and critical remarks. 
I am also indepted to J.~D.~Bjorken, who explained to me his intriquing observation about $\eta_c$
decay, and to B.~Buschbeck for relevant literature on clustering in pp collisions.
This work was supported in parts by the US-DOE grant
DE-FG-88ER40388.

\end{document}